\documentclass[]{gMOS2e}

\begin{document}
\doi{10.1080/0892702YYxxxxxxxx}
\issn{1029-0435}  \issnp{0892-7022}
\jvol{00} \jnum{00} \jyear{2011} 

\markboth{L. Figueras and J. Faraudo}{Molecular Simulation}

\articletype{Preliminary Communication}

\title{Competition between hydrogen bonding and electric field in single-file transport of water in carbon nanotubes}

\author{Luis Figueras and Jordi Faraudo \thanks{$^\ast$Corresponding author. Email: jfaraudo@icmab.es
\vspace{6pt}}\\{\vbox{\vspace{12pt}{\em{Institut de Ci\`encia de Materials de Barcelona (ICMAB-CSIC), Campus UAB, E-08193 Bellaterra, Spain.}}\\\vspace{6pt}\received{Received 17 May 2011}}} }

\maketitle

\begin{abstract}
Recent studies have shown the possibility of water transport across carbon nanotubes, even in the case of nanotubes with small diameter (0.822 nm). In this case, water shows subcontinuum transport following an ordered 1D structure stabilized by hydrogen bonds. In this work, we report MD simulations describing the effect of a perpendicular electric field in this single-file water transport in carbon nanotubes. We show that water permeation is substantially reduced for field intensities of 2-3 V/nm and it is no longer possible under perpendicular fields of 4 V/nm.
\end{abstract}

\section{Introduction}

Despite its strong hydrophobic character, single walled carbon nanotubes (SWCNTs) are spontaneously filled by water \cite{Wu2010,Kumar2011}. Transport of water in these systems has novel and promising properties which have been extensively studied by Molecular Dynamics (MD) simulations (see for example \cite{Review} for a review). Quite interestingly, the water transport trough a carbon nanotube depends strongly on its radius and MD simulations predict a transition between continuum to subcontinuum transport as the radius decreases \cite{Thomas2009}, as recently confirmed experimentally \cite{Exps} . In the case of the nanotubes with small radius, the nanotubes are filled by a one-dimensional ordered chain of water molecules which maintain 2 hydrogen bonds per molecule inside the CNT \cite{Hummer2001}. This effective 1D water system has extremely interesting properties and it has been studied experimentally \cite{Exps}, by MD simulations \cite{Hummer2001,Thomas2009,Wu2010,Review,Zuo2010,Kumar2011,Su2011} and analytical treatments based on the 1D Ising model \cite{Kofinger2010}. Interestingly, this system has been analyzed as a model for water transport in biological pores \cite{Schulten2003}. 

A recent MD simulation study has shown the strong response of this single-file water system inside CNTs to electric fields parallel to the axis of the nanotube \cite{Su2011}, due to the peculiarities of the 1D hydrogen bonding network. Motivated by this previous study, in this work we consider a related but different question, namely the effect of applying an electric field perpendicular to the single file water transport across carbon nanotubes. In this case, a competition arises between the hydrogen bonding network (which tends to maintain the alignment of the chain of water molecules in the nanotube axis direction) and the perpendicular electric field (which tends to align the water dipole in the perpendicular direction). A simple numerical calculation proposed in \cite{Faraudo} suggests that this effect will appear for field intensities between 2-3 V/nm. Noting that a single water-water hydrogen bond has a free energy of about 5$k_BT$ and the dipole of the water molecule in liquid phase is about 2.5 D we can say that H-bonding will compete with external fields with intensities up to $E\sim 2.4 $ V/nm. Albeit high, these field intensities can be studied in simulations and experiments. In this work, we report a preliminary set of MD simulations in order to study the effect on water transport of this competition between hydrogen bonding and electric field.

\section{Methods}

We have performed a total of 33 molecular dynamics (MD) simulations of water transport across carbon nanotubes in both equilibrium and nonequilibrium conditions in presence and in absence of an electric field perpendicular to the axis of the nanotubes. 

All simulations described here were performed using the 2.7 version of the NAMD program \cite{NAMD}. The preparation of the system and analysis of the results were made using the Visual Molecular Dynamics (VMD) software \cite{VMD}. 
\begin{figure}[htp]
\begin{center}
 \includegraphics*[width=8cm]{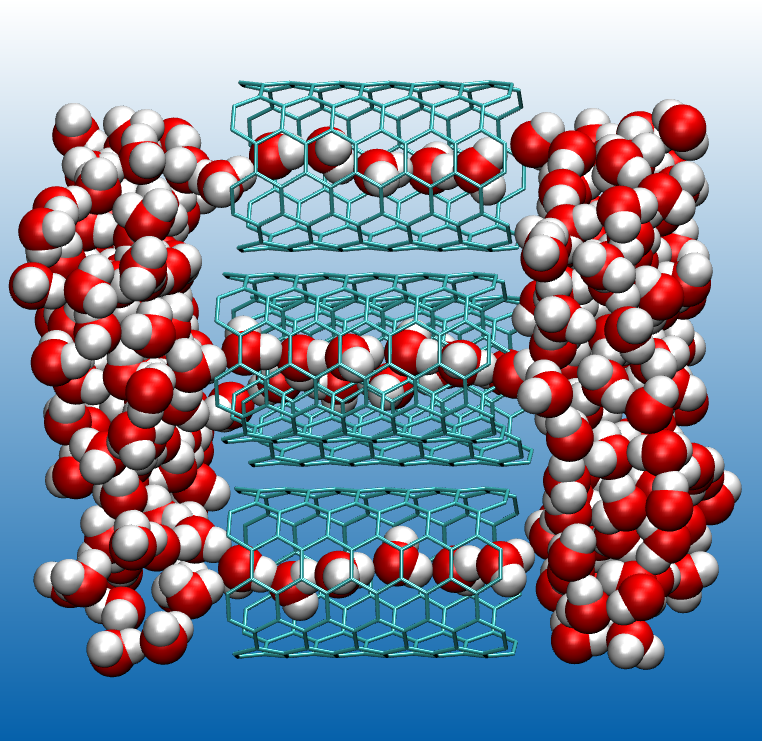}
\caption{Representative snapshot of MD simulations. Atoms from water molecules are shown with Van der Waals radius and carbon atoms from the four SWCNT are shown with lines. Note that water molecules fill the tubes as single file with a strong orientation.}
      \label{Fig:system}
   \end{center}
 \end{figure}
The simulated system is shown in Figure \ref{Fig:system}. It consisted of a periodic simulation box of dimensions $23.0 \times 19.9 \times 30.4 $ \AA$^3$ containing 248 water molecules and a porous membrane made of four carbon nanotubes with their axis aligned in the $z$ direction. Each carbon nanotubes has 144 carbon atoms and has a radius of 0.411 nm and length 1.34 nm.  Water molecules were modelled using the TIP3P model as implemented in the CHARMM force field and carbon atoms from nanotubes were modelled as Lennard-Jones spheres with parameters $\epsilon = 0.07$ kcal/mol and $\sigma=3.9848$\AA\ . Carbon atoms were maintained fixed during all the simulation as in previous works\cite{Hummer2001,Thomas2009}. Lennard-Jones interactions were computed using a smooth (10-12 \AA) cutoff, as it is customary done in NAMD2 simulations. The electrostatic interactions were calculated using the particle-mesh Ewald (PME) method. We employ a multiple time step. The equations of motion are solved with a 1 fs time step, nonbonded interactions are updated each 2 time steps and electrostatic interactions are updated each 4 time steps. The temperature was maintained at 300 K in all simulations employing a Langevin thermostat with a 5 ps$^{-1}$ dumping constant. The Langevin thermostat is applied to the water oxygen atoms only.

Some of our simulations were performed in presence of an external pressure difference. This external pressure was applied in NAMD2 as a Tcl force (as implemented in \cite{NAMDtutorial}) which acts on oxygen atoms in a slab of 5.4 \AA\ thickness (all oxygen atoms with coordinates $z>12.5$ \AA\ or $z<-12.5$ \AA). This methodology is also described in detail in Ref \cite{Schulten2001}. The applied pressure $\Delta P$ is computed from the force $f$ by $\Delta P = nf/A$ where $n$ is the number of water molecules in the slab and $A$ its transversal area.

In most of our simulation runs, we applied an uniform electric field $E_y$, perpendicular to the axis of the nanotubes, a feature which is implemented in NAMD2. The field is acting in all the simulation box.

In all simulations, we compute the number of permeation events, defined as the number of water molecules observed to completely cross the nanotube (entering from one end and leaving the tube from the other end). Obviously this quantity depends on the simulation time, so it was normalized to the total production time (in ns). The length of the production runs was decided by monitoring this quantity, i.e. simulations were performed until this quantity reached an stable average value. Depending on the particular simulation runs, simulation times are between 5 and 40 ns. 

\section{Results and Discussion}

 First, we conducted equilibrium simulations (no applied external fields or pressure gradients). In this case, the average flux of water is negligible, as expected in equilibrium. In equilibrium we obtain an average rate of permeation events of water molecules of 9 molecules/ns in any of the directions. 

\begin{figure}[htp]
\begin{center}
 \includegraphics*[width=9cm]{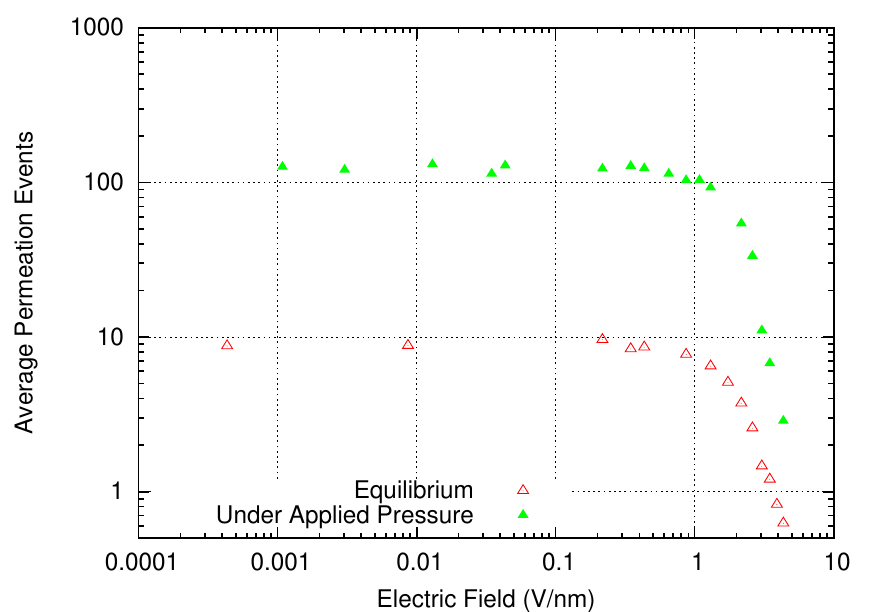}
\caption{Permeation events of water molecules per ns observed in MD simulations as a function of an applied electrical field perpendicular to the axis of the nanotubes. Open triangles correspond to equilibrium simulations and permeation events are averaged over the $+z$ and $-z$ direction. Filled triangles correspond to simulations with a water flow induced under an external pressure of $\Delta P = 3714$ atm. In this case, permeation events occur only in the direction of decreasing pressure.}
      \label{Fig:results}
   \end{center}
 \end{figure}
\begin{figure}[htp]
\begin{center}
 \includegraphics*[width=9cm]{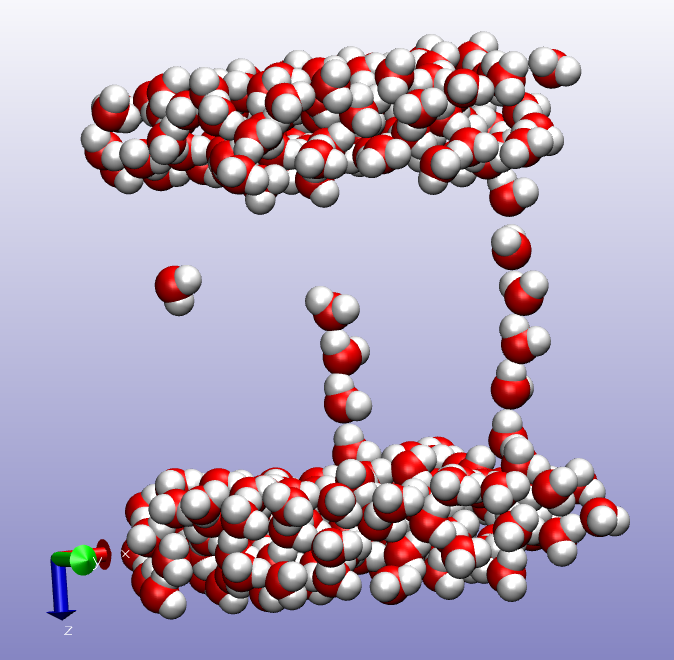}
\caption{Snapshot of MD simulations under an electric field of 2 V/nm in the $y$ direction. Atoms from water molecules are shown with Van der Waals radius and carbon atoms from the four SWCNT are not shown for simplicity. Note that one of the tubes is empty and in another nanotube there is only a single, isolated water molecule inside one of the nanotubes oriented in the field direction. In another of the tubes the hydrogen-bonded water chain is still intact and water molecules fill the tube.}
      \label{Fig:dehydration}
   \end{center}
 \end{figure}

Now, we consider the effect of applying an external electric field $E_y$ perpendicular to the axis of the nanotubes. The results are shown in Figure  \ref{Fig:results}. Fields up to 1 V/nm have no significant impact on water transport. However, for larger fields, the number of permeation events decreases substantially as the field strength increases. For fields about 2 V/nm, the number of permeation events has been reduced to less than its equilibrium value (see Fig. \ref{Fig:results}). We also observe that chains inside nanotubes start to break and we even observe crossing of isolated (not hydrogen bonded) water molecules across the nanotube, as shown in the snapshot in Figure \ref{Fig:dehydration}. Larger fields reduce the permeation events even more and it can be say that water permeation is not possible for fields larger than 4-5 V/nm.

We have also performed simulations under a large external pressure $\Delta P=3714$ atm (see Fig.\ref{Fig:results} ). In absence of external electrical field, we observe about 133 permeation events per ns. In all cases permeation occurs in the direction of decreasing pressure, as should be expected in this case. Under the application of external, perpendicular electric field we obtain results strongly resembling to those obtained in the absence of a pressure gradient. Again, the effect of fields smaller than 1 V/nm is negligible, and substantial reduction of the flow of water requires fields of the order of 2-3 V/nm. It can be said that the presence of the external pressure has no effect on the competition between hydrogen bonding and the external electric field. 

\section{Conclusions}

In this work, we have performed MD simulations of single-file transport of water in carbon nanotubes of small radius. Simple arguments suggest that the application of an electric field perpendicular to the axis of the carbon nanotube will disrupt the hydrogen bond structure of this 1D water system and affect the transport properties of water inside the nanotube. Our simulations show that water transport is not affected under perpendicular fields up to 1 V/nm. Larger fields substantially reduce the flow of water, and in fact transport is not possible for perpendicular fields above $4$ V/nm. Our work shows the possibility of using perpendicular electric fields as nanofluidic ``switches'' to regulate water transport in carbon nanotubes. It is also interesting to note that a previous work \cite{Su2011} has shown the advantages of pumping water across nanotubes using parallel electric fields (between 0.01 V/nm - 1 V/nm) instead of using a pressure gradient. It could be interesting to analyze also this parallel case with higher fields (2-4 V/nm) which could also somewhat influence the hydrogen bonding network. Our own ongoing work indicates that under parallel fields $\sim$ 4 V/nm there are significant structural changes in the single file of water molecules. Overall, all these results show the interesting possibilities arising from using electric fields to control water flow in carbon nanotubes.

\section*{Acknowledgments}
This work is supported by the Spanish Government (grants FIS2009-13370-C02-02 and CONSOLIDER-NANOSELECT-CSD2007-00041) and Generalitat de Catalunya (grant 2009SGR164).

\end{document}